\documentclass[universe,article,accept,pdflatex,moreauthors]{Definitions/mdpi} 
\usepackage[utf8]{inputenc}
\usepackage{enumitem}
\usepackage{graphicx}
\usepackage{amscd}
\usepackage{slashed}
\usepackage{amssymb}
\usepackage{mathtools} 
\usepackage{pstricks}
\usepackage[mathscr]{eucal}
\usepackage{bbm}
\usepackage{xcolor}
\usepackage{url}

\firstpage{1} 
\pubvolume{1}
\issuenum{1}
\articlenumber{0}
\pubyear{2025}
\copyrightyear{2025}
\externaleditor{Firstname Lastname} % More than 1 editor, please add `` and '' before the last editor name
\datereceived{4 March 2025} 
\daterevised{11 April 2025} % Comment out if no revised date
\dateaccepted{24 April 2025} 
\datepublished{ } 
\hreflink{https://doi.org/} % If needed use \linebreak
\Title{Time-like Extra Dimensions: Quantum Nonlocality, Spin, and Tsirelson Bound}

\TitleCitation{Time-like Extra Dimensions: Quantum Nonlocality, Spin, and Tsirelson Bound}

\Author{{Mohammad Furquan} %MDPI: Please carefully check the accuracy of names and affiliations.
 $^{1}$, {Tejinder} %MDPI: To AE: the format of this name in tex is correct. Please change the name on susy to make them consistent.
 {P.} %MDPI: We changed lowercase p to uppercase. Please confirm.
 Singh $^{2,}${*} %MDPI: We added the symbol that represents the corresponding author here according to that submitted online at susy. Please confirm.
 and {P} %MDPI: We added a dot for the abbreviated name. Please check and confirm.
 Samuel Wesley $^{3}$}

\AuthorNames{Mohammad Furquan, Tejinder P. Singh and P Samuel Wesley}

\isAPAStyle{%
       \AuthorCitation{Lastname, F., Lastname, F., \& Lastname, F.}
         }{%
        \isChicagoStyle{%
        \AuthorCitation{Lastname, Firstname, Firstname Lastname, and Firstname Lastname.}
        }{
        \AuthorCitation{Furquan, M.; Singh, T.P.; Wesley, P.S.}
        }
}

\address{%
$^{1}$ \quad Department of Physics, Indian Institute of Technology, Delhi 110016, India; {mohammad.furquan.ph1210208@physics.iitd.ac.in} %MDPI: We changed uppercase letters in this email to lowercase. Please confirm. This email is different from that (ph1210208@iitd.ac.in) on susy. Please confirm which one is correct.
\\
$^{2}$ \quad Tata Institute of Fundamental Research, Homi Bhabha Road, Mumbai 400005, India\\
$^{3}$ \quad Department of Theoretical Physics, University of Madras, Chennai 600025, India; samwesprem7@gmail.com}

\corres{Correspondence: tpsingh@tifr.res.in}

\abstract{The $E_8 \otimes  E_8$ octonionic theory of unification suggests that our universe is six-dimensional and that the two extra dimensions are time-like. These time-like extra dimensions, in principle, offer an explanation of the quantum nonlocality puzzle, also known as the EPR paradox. Quantum systems access all six dimensions, whereas classical systems such as detectors experience only four dimensions. Therefore, correlated quantum events that are time-like separated in 6D can appear to be space-like separated and, hence, nonlocal, when projected to 4D. Our lack of awareness of the extra time-like dimensions creates the illusion of nonlocality, whereas, in reality, the communication obeys special relativity and is local. Bell inequalities continue to be violated because quantum correlations continue to hold.  In principle, this idea can be tested experimentally. We develop our analysis after first constructing the Dirac equation in 6D using quaternions and using the equation to derive spin matrices in 6D and then in 4D. We also show that the Tsirelson bound of the CHSH inequality can in principle be violated in 6D.}


\begin{document}

%%%%%%%%%%%%%%%%%%%%%%%%%%%%%%%%%%%%%%%%%%
    \section{Introduction}
    The EPR paradox notes the following property of quantum mechanics: Given a correlated pair of, say, two quantum subsystems, a measurement on one subsystem influences the other subsystem nonlocally (note that causality is not violated). This property has been confirmed by experiments and is equivalent to the confirmation that quantum systems violate Bell's inequalities. Such a Bell-type measurement cannot be used to transmit information faster than light; therefore, the laws of special relativity are not violated, in spite of there being a nonlocal influence. One could accept this peculiarity as an inevitable feature of quantum mechanics and assert that the collapse of the wave function is accompanied by a nonlocal effect and that there is nothing more to be explained.  Alternatively, one could insist that the following needs to be explained: What is the physical mechanism whereby subsystems in a correlated quantum system impact each other outside the light cone \cite{gisinl, Maudlin}?  Could it be that our understanding of spacetime structure in quantum theory is incomplete? It is this latter stance that is adopted in the present brief article, and a simple solution is proposed, which removes this tension between quantum mechanics and special relativity, without altering the laws of either theory.

    Our proposal is that the spacetime of our universe is not four-dimensional but six-dimensional and that the two extra dimensions are time-like. Laws of special relativity also hold in this bigger spacetime, with the signature $(3,3)$ and the influence of wave function collapse, which takes place locally in 6D spacetime. Correlated events that are time-like separated in 6D can appear to be space-like separated in 4D, giving rise to the EPR paradox. Bell's inequalities continue to be violated, in 6D as well as in 4D. This is so because the violation is equivalent to ruling out a deterministic locality. Indeterministic locality is permitted (which is what gives rise to higher-than-classical quantum correlations). In 4D, this translates into an apparent indeterministic nonlocality, giving rise to the illusion of a conflict between quantum mechanics and relativity.

    In the next section, we briefly introduce the Dirac equation  in 6D spacetime. Also, one can embed two 4D spacetime manifolds in 6D, having relatively flipped signatures. Classical systems, including detectors, live in only one of the two 4D submanifolds.  
     In Section \ref{sec3}, we use the Dirac equation to derive spin matrices in 6D and then in 4D. 
    In Section \ref{sec4}, we explain the resolution of the EPR paradox in some detail and comment on the hypothetical possibility of its experimental validation. In Section \ref{sec5}, we describe the motivation for extra time-like dimensions coming from our ongoing research program of gravi-weak unification.  In Section \ref{sec6}, we note that the Tsirelson bound in the CHSH inequality can sometimes be violated.
   
  {We introduce the concept of absolute time in special relativity and assume that in a Bell-type experiment, the influence of wave function collapse is transmitted (causally) by a mediating field. We emphasize to the reader that our proposed new theory deviates fundamentally from standard quantum mechanics (QM) and standard special relativity theory (SRT) while agreeing with their present experimental predictions. In standard QM, there is no field carrying information from Alice to Bob, and, in SRT, there is no absolute time (even though empirically equivalent theories can be constructed that do have an absolute notion of time, but these are not SRT but different theories, like Lorentz's ether theory)}.
   
    {In Section \ref{sec7} (Critique), we carefully examine the underlying assumptions on which our proposal is based and provide justification for those assumptions.} 

\section{Some Remarks on Dirac Equation in 6D Spacetime \label{sec2}}
\subsection{Dirac Equation Using Gamma Matrices}

We start with the 6D spacetime manifold \(M_6\) with the coordinates \((t_1,t_2,t_3,x_1,x_2,x_3)\) and signature $(+, +, +, -, -, -)$. %MDPI: Please confirm if hyphen should be minus sign $-$. If so, please check full text and revise.
 The Minkowski metric with the isometry group $SO(3,3)$ is
\begin{equation}
    ds^2 = dt_1^2 + dt_2^2 + dt_3^2 - dx_1^2 - dx_2^2 - dx_3^2
\end{equation}
{The trajectory} %MDPI: Please confirm whether the paragraphs starting with capital letters under the formula should be indented. If so, please check full text and revise.
 of a particle in \(M_6\) can be identified by the functions \(t^i(\tau),x^i(\tau)\), where \(\tau \) is the proper time in \(M_6\). The Klein--Gordon equation is
\begin{equation}
(\square +m^2)\psi=0 \quad ; \quad \square= \partial_{t1}^2+\partial_{t2}^2+\partial_{t3}^2-\partial_{x1}^2-\partial_{x2}^2-\partial_{x3}^2
\end{equation}

We assume that prior to the electroweak symmetry breaking, the universe is endowed with a 6D spacetime. The (chiral) symmetry breaking gives rise to two overlapping 4D spacetimes, one of which we are familiar with and whose pseudo-Riemannian geometry is determined by the laws of general relativity. The other 4D spacetime has a signature flipped with respect to ours, and its pseudo-Riemannian geometry can be shown to be a gravity-type interpretation of the weak force. The two spacetimes have one time and one space dimension in common. The two time dimensions of the other spacetime that are not part of this intersection can be interpreted as the internal symmetry directions overlaid (as a vector bundle) on our 4D spacetime. Prior to the symmetry breaking, gravitation and the electroweak interaction are unified into a 6D (non-chiral) theory of gravitation \cite{Singh2024Review}.

The following resolution of the EPR paradox depends only on there being a 6D spacetime of signature (3,3). It is independent of the above theoretical motivation for the 6D spacetime, coming from our approach to gravi-weak unification.

Inside the 6D spacetime, the 4D submanifold \(M_4\) with symmetry $SO(1,3)$ has the coordinates \(
(t_1,x_1,x_2,x_3)\) with the signature $(+,-,-,-)$. A second 4D submanifold \(M_4'\) with symmetry $SO(3,1)$  can be chosen to be \((x_1,t_1,t_2,t_3)\), which has the signature $(-,+,+,+)$. Instead of choosing \(x_1\), we could also choose $x$, which is any linear combination of \(x_1,x_2\), and \(x_3\). This shows that there is a common plane $(t_1,x)$ between these manifolds. We should get the 6D Dirac equation using 
\begin{equation}
\square=(\Gamma^\mu\partial_{\mu})(\Gamma^\nu\partial_{\nu}) 
\end{equation}
with \(\{\Gamma^\mu,\Gamma^\nu\}=2\eta^{\mu\nu}I\). In the 4D case, imposing the anti-commutation conditions on the four gamma matrices, we obtain a $4\times4$ matrix representation. What is the possible representation in this case of six gamma matrices with \((\Gamma^{i})^2=I\) for $i=1,2,3$ and \((\Gamma^{i})^2=-I\) for $i=4,5,6$? Using the Clifford algebra representation, we obtain spinors of dimension \(2^{6/2}=8\) and a set of $8\times 8$ gamma matrices. These gamma matrices can be constructed explicitly by choosing representations, and we can also construct the corresponding Lagrangian. This gives us a Dirac equation \((i\Gamma^\mu\partial_\mu-Q)\psi=0\) for \(M_6\), where Q is a general source charge, and this equation should also reduce to the Dirac equation in \(M_4\), as we will see later. The 8D spinors in \(M_6\), as we will explain, will break into two Dirac spinors, one belonging to the section $M_4$ and the other belonging to the section $M_4'$. Thus, we will be able to write
\begin{equation}
\psi_{M_6}=\psi_{M_4}+\psi_{M_4'}
\end{equation}
This will allow us to understand that the quantum state is a superposition across these two submanifolds. We also need to define a new Hilbert space \(H \oplus  H'\). We need to include more observables in our quantum system, a set of observables \(\{A\}\) in $H$, a mirror set of observables \(\{A'\}\) in \(H'\), and some additional observables forming a complete set of observables for \(H \oplus H'\):
%Because the new Hilbert space is just the direct sum, %Tsirelson bound will remain the same since the original set %of observables doesn't act on states corresponding to \
%(M_4'\).
\begin{align}
\label{blockrep}
    \left(
    \begin{array}{c|c}
      \{ A\} & \\
      \hline
       & \{A'\}
    \end{array}
    \right)
    \begin{pmatrix}
        \psi_{M_{4}} \\
        \psi_{M'_{4}}
    \end{pmatrix}
\end{align}
This describes the actions of different observables on the system in \(M_6\), which is a superposition across \(M_4\) and \(M_4'\). These off-block-diagonal terms will vanish only if the set of \(\psi_{M_4}\) and \(\psi_{M_4'}\) forms an invariant subspace under operators that are defined on \(H\oplus H'\). These are the operators that come from the direct sum of space of linear operators \(\{A\}\) and \(\{A'\}\). However, in this larger Hilbert space, it is possible to have operators for which  \(\psi_{M_4}\) and \(\psi_{M_4'}\) do not form an invariant subspace and these off-block-diagonal terms do not vanish. In the context of gravi-weak unification, this could suggest a mixing between these interactions. We will see later in Section \ref{sec6} that such operators can also lead to a possible violation of the Tsirelson Bound. We want to investigate such operators and interpret this mixing. In Section \ref{sec3}, we explicitly show how to construct the spin operators in \(H\) and \(H'\).
\subsection{Dirac Operator Using Quaternions}
We propose an alternative construction of the Dirac operator using quaternions. To construct the 6D Dirac operator, we start with the algebra of split biquaternions, expressed as   \(\mathbb{H}+\omega\mathbb{H}\), with \(\omega\) satisfying \(\tilde{\omega}=-\omega\), \(\omega^2=1\). We have pure imaginary quaternions associated with each set of quaternions as \((\hat{i}, \hat{j}, \hat{k}\)) and (\(\hat{l}, \hat{m}, \hat{n}\)). The elements inside each set anticommute with each other, and elements of one set commute with the elements of the other set. These can be used as vectors in 3D spaces, and we would like to use them to describe our 6D spacetime. We express the events in 6D as
\begin{equation}
x_6=t_1\hat{i}+t_2\hat{j}+t_3\hat{k}+\omega(x_1\hat{l}+x_2\hat{m}+x_3\hat{n})
\end{equation}
which gives 
\begin{equation}
x_6\tilde{x}_6=t_1^2+t_2^2+t_3^2-x_1^2-x_2^2-x_3^2
\end{equation}
This is the correct signature for the interval and is associated with SO(3,3). Note that this \(x_6\) is not Hermitian. Let us denote \(t_1, t_2, t_3, x_1, x_2, x_3\) by the indices 01, 02, 03, 1, 2, 3. Now, we  construct the Dirac operator as
\begin{equation}
D_6=\hat{i}\partial_{01}+\hat{j}\partial_{02}+\hat{k}\partial_{03}+\omega(\hat{l}\partial_1+\hat{m}\partial_2+\hat{n}\partial_3)
\end{equation}
which gives us the correct Klein--Gordon operator
\begin{equation}
D_6\tilde{D}_6=\partial_{01}^2+\partial_{02}^2+\partial_{03}^2-\partial_1^2-\partial_2^2-\partial_3^2
\end{equation}
Note that squaring the Dirac operator will not work as it will give additional terms.
We then ask for a Dirac equation with a general source charge:
\begin{equation}
    ihD_6\psi=Q\psi
\end{equation}
This is similar to the one obtained above, but now we have quaternions instead of gamma matrices. For relating this to 4D spacetimes, this 6D Dirac operator can be broken down into two Dirac operators \(D_4\) and \(D_4'\):
\begin{equation}
D_4=\hat{i}\partial_{01}+\omega(\hat{l}\partial_1+\hat{m}\partial_2+\hat{n}\partial_3)
\end{equation}
which gives
\begin{equation}
D_4\tilde{D}_4=\partial_{01}^2-\partial_1^2-\partial_2^2-\partial_3^2
\end{equation}
Similarly, for the Dirac operator, \(D_4'\) we can write 
\begin{equation}
D_4'=\omega \hat{l}\partial_1+\hat{i}\partial_{01}+\hat{j}\partial_{02}+\hat{k}\partial_{03}
\end{equation}
This also gives the correct signature. Both these Dirac operators give the Dirac equation for \(M_4\) and \(M_4'\). We also note that Wilson \cite{Wilson} showed, using \(SO(3,3)\!\!\sim\!\!SL(4,\mathbb{R})\)  and the generators of \(SL(4,\mathbb{R})\) with two copies of 4D spacetimes described in terms of gamma matrices and quaternions, that it is easy to make maps between the quaternion vectors $\hat i, \hat j, \hat k, \hat l, \hat m, \hat n$ and the quaternion bivectors to gamma matrices. We use this to reduce this quaternionic Dirac operator to the usual one. Use \(\hat{i} \mapsto \gamma^0,\omega\hat{l} \mapsto\gamma^1,\omega\hat{m}\mapsto\gamma^2,\omega\hat{n}\mapsto\gamma^3\) to obtain the usual Dirac~equation. 

Alternatively, we can also construct the 4D Dirac operator as 
\begin{equation}
D_{4}=\partial_0-i(\hat{i}\partial_1+\hat{j}\partial_2+\hat{k}\partial_3)
\end{equation}
This satisfies the above equation, and it is Hermitian. The Hermiticity is not preserved in the other cases mentioned above. 
\subsection{Dirac Equation Using Quaternion Bivectors}
Another attempt motivated by Lambek's work {\cite{rf3lambek3}} %MDPI: We rearranged the references to make them appear in numerical order. Please confirm.
 was to define \(x_6\) using the bivectors. We have nine bivectors: \(\hat{i}\hat{l},\ \hat{j}\hat{l},\ \hat{k}\hat{l}\), \( \hat{i}\hat{m},\ \hat{j}\hat{m},\ \hat{k}\hat{m},\) \(\hat{i}\hat{n},\ \hat{j}\hat{n},\ \hat{k}\hat{n}\). In this case, the constructed \(x_6\) will only use six of these bivectors and the Dirac operator will be written as \begin{equation}
    D_6=\hat{l}\hat{i}\partial_{01}+\hat{l}\hat{j}\partial_{02}+\hat{l}\hat{k}\partial_{03}+\omega(\hat{m}\hat{i}\partial_1+\hat{m}\hat{j}\partial_2+\hat{m}\hat{k}\partial_3)
\end{equation}
which gives
\begin{equation}
\begin{aligned}
    D_6 \tilde{D}_6 &= \partial_{01}^2 + \partial_{02}^2 + \partial_{03}^2 
    - \partial_{1}^2 - \partial_{2}^2 - \partial_{3}^2 \\
    &\quad - 2\omega \hat{l}\hat{m}\hat{i}\hat{j} \partial_{01}\partial_{2} 
    - 2\omega \hat{l}\hat{m}\hat{i}\hat{k} \partial_{01}\partial_{3} \\
    &\quad - 2\omega \hat{l}\hat{m}\hat{j}\hat{i} \partial_{02}\partial_{1} 
    - 2\omega \hat{l}\hat{m}\hat{j}\hat{k} \partial_{02}\partial_{3} \\
    &\quad - 2\omega \hat{l}\hat{m}\hat{k}\hat{i} \partial_{03}\partial_{1} 
    - 2\omega \hat{l}\hat{m}\hat{k}\hat{j} \partial_{03}\partial_{2}.
\end{aligned}
\end{equation}

This is not giving the correct signature, and one has these extra terms. They cancel when we impose the constraints \begin{equation}
   \partial_{01}\partial_2=\partial_{02}\partial_1,\quad \partial_{01}\partial_3=\partial_{03}\partial_1, \quad \partial_{02}\partial_3=\partial_{03}\partial_2 
\end{equation}
How do we interpret these odd relations?
Based on Wilson's work \cite{Wilson},  this can also be related to gamma matrices as (\(\hat{l}\hat{i},\hat{l}\hat{j},\hat{l}\hat{k}\)) correspond to (\(\gamma_0\gamma_1, \gamma_0\gamma_2, \gamma_0\gamma_3\)) and (\(\hat{m}\hat{i},\hat{m}\hat{j},\hat{m}\hat{k}\)) correspond to (\(i\gamma_1,i\gamma_2,i\gamma_3\)). This means that \begin{equation}
    D_6=\gamma_0(\gamma_1\partial_{01}+\gamma_2\partial_{02}+\gamma_3\partial_{03})+\omega(\gamma_1\partial_1+\gamma_2\partial_2+\gamma_3\partial_3)
\end{equation}
Since \(\gamma_0\gamma_1\) commutes with \(\gamma_1\) but not with \(\gamma_2\), we obtain similar extra terms while computing \(D_6\tilde{D_6}\). This is fixed by mapping \(\omega\gamma_i\) to some new gamma matrices that complete the set of six gamma matrices(\(\Gamma^i, \tilde\Gamma^i\)) for 6D.
Let us also try to construct \(D_4\) and \(D_4'\) from this: \begin{equation}
    D_4= \hat{l}\hat{i}\partial_{01}+\omega(\hat{m}\hat{i}\partial_1+\hat{m}\hat{j}\partial_2+\hat{m}\hat{k}\partial_3)
\end{equation}
This shows that\begin{equation}
  D_4\tilde{D_4}=\partial_{01}^2-\partial_1^2-\partial_2^2-\partial_3^2-2\omega\hat{l}\hat{m}\hat{i}\hat{j}\partial_{01}\partial_{2}-2\omega\hat{l}\hat{m}\hat{i}\hat{k}\partial_{01}\partial_{3}  
\end{equation}
thus, again, giving extra terms. 

We will therefore adhere to our earlier method of deriving the 6D and 4D Dirac equations from quaternions. An analogous decomposition (from 6D to 4D) of the Klein--Gordon equation can be expected to hold for bosons.

\section{Interpreting Spin in 6D \label{sec3}}
In this section, we will explore the interpretation of spin by considering the non-relativistic limit of the Dirac equation in 6D. By examining this limit and projecting to both $M_{4}$ and $M'_{4}$, we aim to identify the spin matrices. This framework can potentially facilitate the understanding of the behavior of spin in the presence of additional time-like dimensions. Therefore, consider the Dirac equation of the form
\begin{align}
    \left(i\Gamma^{\mu} \nabla_{\mu} - Q \right) \Phi = 0
\end{align}
where the set of gamma matrices are
\begin{align}
    [\hat{\Gamma}_{3,3}] = \left\{ \tilde{\Gamma}_{3}, \tilde{\Gamma}_{2}, \tilde{\Gamma}_{1}, \Gamma_{1}, \Gamma_{2}, \Gamma_{3}\right\} \label{nrgamma3,3}
\end{align}
such that
\begin{align}
    \{\Gamma^{\mu}, \Gamma^{\nu} \} = 2 \eta^{\mu \nu} \mathbbm{1}_{8}
\end{align}
where $\eta_{\mu \nu} = \text{diag}(+1,+1,+1,-1,-1,-1)$. 
Here, $\tilde{\Gamma}_{i}^{2}=\mathbbm{1}_{8}$, and $\Gamma_{i}^{2}=\mathbf{i}_{8}^{2}$ for $i=1,2,3$. 
Let us introduce the following decomposition:
\begin{align}
    \tilde{\Gamma}_{i} = i\mathcal{B} \lambda^{i} \hspace{3em} (i=1,2,3)\\
    \Gamma_{i} = \mathcal{B} \alpha^{i} \hspace{3em} (i=1,2,3)
\end{align}
such that
\begin{align}
    \{\alpha_{i},\alpha_{j}\} &= 0 \hspace{2em} (i \neq j)\\
    \{\lambda_{i},\lambda_{j}\} &= 0 \hspace{2em} (i \neq j)\\
    \{\alpha_{i},\lambda_{j}\} &= 0 \hspace{2em} (\forall i, j)\\
    \alpha_{i}^{2} = \lambda_{i}^{2} &= \mathcal{B}^{2} = \mathbbm{1}
\end{align}
Then, the Dirac equation can be written in a {suitable form:} %MDPI: Please remove the border around the equation if it is not necessary.
\begin{align}
    {i\lambda^{i}\tilde{\partial}_{i}\Phi = \left( -i\alpha^{i}\partial_{i} + \mathcal{B} Q \right) \Phi} \label{nrdirac1}
\end{align}
To take the non-relativistic limit, we will choose a basis wherein
\begin{align}
    \alpha^{i} &= \Sigma^{i} \begin{pmatrix}
    0 & \mathbbm{1}  \\
    \mathbbm{1}  & 0
    \end{pmatrix} \\
    \lambda^{i} &= \tilde{\Sigma}^{i} \begin{pmatrix}
    \mathbbm{1} & 0  \\
     0 & \mathbbm{1}
    \end{pmatrix} \\
    \mathcal{B} &= \begin{pmatrix}
    \mathbbm{1} & 0 \\
    0 & -\mathbbm{1}
    \end{pmatrix} \\
    [\tilde{\Sigma}_{i}, \Sigma_{j}] &= 0 \hspace{2em} (\forall i, j)
\end{align}
and decompose $\Phi$ as
\begin{align}
    \Phi = \begin{pmatrix}
    \phi_{1}  \\
    \phi_{2}
    \end{pmatrix} = \begin{pmatrix}
    \varphi_{1}  \\
    \varphi_{2} \\
    \varphi_{3} \\
    \varphi_{4}
    \end{pmatrix}\label{6dspinordecomp1}
\end{align}
where both $\phi_{1}$ and $\phi_{2}$ are Dirac spinors.
Substituting (\ref{6dspinordecomp1}) into (\ref{nrdirac1}), we obtain
\begin{align}
    i\begin{pmatrix}
    \tilde{\Sigma}^{i}\tilde{\partial}_{i} & 0 \\
    0 & \tilde{\Sigma}^{i}\tilde{\partial}_{i}
    \end{pmatrix} \begin{pmatrix}
    \phi_{1}  \\
    \phi_{2}
    \end{pmatrix} = \left[-i\begin{pmatrix}
    0 & \Sigma^{i}\partial_{i} \\
    \Sigma^{i}\partial_{i} & 0
    \end{pmatrix} + \begin{pmatrix}
    \mathbbm{1} & 0 \\
    0 & -\mathbbm{1}
    \end{pmatrix} Q \right] \begin{pmatrix}
    \phi_{1}  \\
    \phi_{2}
    \end{pmatrix}
\end{align}
The source charge matrix $Q$ in our choice of bases is
\begin{align}
    Q = \begin{pmatrix}
    m & 0 & 0 & 0\\
    0 & 0 & 0 & m'\\
    0 & 0 & m & 0\\
    0 & m' & 0 & 0
    \end{pmatrix}
\end{align}
This results in a system of four coupled differential equations:
\begin{align}
    i(\tilde{\Sigma}^{i}\tilde{\partial}_{i})\varphi_{1} = -i(\Sigma^{i}\partial_{i})\varphi_{3} + m \varphi_{1} \label{cde1}\\
    i(\tilde{\Sigma}^{i}\tilde{\partial}_{i})\varphi_{2} = -i(\Sigma^{i}\partial_{i})\varphi_{4} + m' \varphi_{4} \label{cde2}\\
    i(\tilde{\Sigma}^{i}\tilde{\partial}_{i})\varphi_{3} = -i(\Sigma^{i}\partial_{i})\varphi_{1} - m \varphi_{3} \label{cde3}\\
    i(\tilde{\Sigma}^{i}\tilde{\partial}_{i})\varphi_{4} = -i(\Sigma^{i}\partial_{i})\varphi_{2} - m' \varphi_{2} \label{cde4}
\end{align}
where $[\tilde{\Sigma}_{i}, \tilde{\Sigma}_{j}] = i\epsilon_{ijk} \hspace{0.2em} \tilde{\Sigma}_{k}$ and $[\Sigma_{i}, \Sigma_{j}] = i\epsilon_{ijk} \hspace{0.2em} \Sigma_{k}$.

\subsection{Non-Relativistic Limit (v/c) Expansion for the First Dirac Spinor ($\mathcal{M}_{4}$)}
Consider the Dirac equation
\begin{align}
    i\lambda^{i}\tilde{\partial}_{i}\Phi = \left( -i\alpha^{i}\partial_{i} + eA^{0}+ \mathcal{B} Q \right) \Phi 
\end{align}
Let us now consider
\begin{align}
    i(\tilde{\Sigma}^{i}\tilde{\partial}_{i})\varphi_{1} = -i(\Sigma^{i}\partial_{i})\varphi_{3} + eA^{0}\varphi_{1} + m \varphi_{1} \\
    i(\tilde{\Sigma}^{i}\tilde{\partial}_{i})\varphi_{3} = -i(\Sigma^{i}\partial_{i})\varphi_{1} + eA^{0}\varphi_{3} - m \varphi_{3}
\end{align}
and decompose $\varphi_{1}$ and $\varphi_{3}$ as follows to describe them as slowly varying functions:
\begin{align}
    \varphi_{1} \rightarrow e^{-imt^{i}} \varphi_{1} \\
    \varphi_{3} \rightarrow e^{-imt^{i}} \varphi_{3}
\end{align}
Substituting these into (\ref{cde1}) and (\ref{cde3}) gives us
\begin{align}
    \tilde{\Sigma}^{i}(m+i\tilde{\partial}_{i})\varphi_{1}=-i(\Sigma^{i}\partial_{i})\varphi_{3} + eA^{0} \varphi_{1}+ m \varphi_{1} \label{cdevc1}\\
    \tilde{\Sigma}^{i}(m+i\tilde{\partial}_{i})\varphi_{3} = -i(\Sigma^{i}\partial_{i})\varphi_{1} + eA^{0}\varphi_{3} - m \varphi_{3} \label{cdevc2}
\end{align}
In the non-relativistic limit, $m\gg eA^{0}$, we have
\begin{align}
    eA^{0} \approx 0 \\
    \tilde{\Sigma}^{i}\tilde{\partial}_{i} \approx 0
\end{align}
So, (\ref{cdevc2}) becomes
\begin{align}
    (\tilde{\Sigma}^{i}m + m)\varphi_{3} = -i(\Sigma^{i}\partial_{i})\varphi_{1} \label{cdevc3}
\end{align}
Substituting (\ref{cdevc3}) into (\ref{cdevc1}), we obtain %MDPI: Please remove the border around the equation if it is not necessary.
\begin{align}
    {\tilde{\Sigma}^{i}(m+i\tilde{\partial}_{i})\varphi_{1}=-(\Sigma^{i}\partial_{i})^{2}(\tilde{\Sigma}^{i}m + m)^{-1}\varphi_{1} + eA^{0} \varphi_{1} + m \varphi_{1}} \label{1paulieqn6d}
\end{align}
Projecting to the spacetime $\mathcal{M}_{4}$ with signature $\eta_{\mu \nu} = \text{diag}(+1,-1,-1,-1)$, we impose \mbox{the following:}
\begin{align}
    \tilde{\Sigma}^{i} \rightarrow \mathbbm{1} \\
    i\tilde{\partial}_{i} \rightarrow i\partial_{t} 
\end{align}
Hence, (\ref{1paulieqn6d}) becomes
\begin{align}
    \mathbbm{1}(m+i\partial_{t})\varphi_{1}&=-(\Sigma^{i}\partial_{i})^{2}(\mathbbm{1}m + m)^{-1} \varphi_{1}+ eA^{0}\varphi_{1} + m \varphi_{1} \\
    i\partial_{t} &=-\frac{1}{2m}(\Sigma^{i}\partial_{i})^{2} + eA^{0} \label{pauli1st}
\end{align}
We thus recover Pauli's equation, which is typically represented as
\begin{align}
    i\partial_{t} = \frac{1}{2m}(\sigma \cdot \hat{p})^{2} + q\phi \label{paulieqn}
\end{align}
Comparing (\ref{pauli1st}) and (\ref{paulieqn}), we see that $\Sigma_{i}$ are the spatial spin matrices.

\subsection{Non-Relativistic Limit (v/c) Expansion for the Other Dirac Spinor ($\mathcal{M}'_{4}$)}
The source charge term \(m'\) should be \(e^2\), and the coupling factor is \(\sqrt{m}\) instead of e (we explain the reason for this in Section \ref{sec5}). Following along with the previous section, let us now consider
\begin{align}
    i(\tilde{\Sigma}^{i}\tilde{\partial}_{i})\varphi_{2} = -i(\Sigma^{i}\partial_{i})\varphi_{4} + \sqrt{m}\tilde{A}^{0}\varphi_{4} + e^{2} \varphi_{4} \label{cde2t}\\
    i(\tilde{\Sigma}^{i}\tilde{\partial}_{i})\varphi_{4} = -i(\Sigma^{i}\partial_{i})\varphi_{2} + \sqrt{m}\tilde{A}^{0}\varphi_{2} - e^{2} \varphi_{2} \label{cde4t}
\end{align}
Decompose $\varphi_{2}$ and $\varphi_{4}$ as follows:
\begin{align}
    \varphi_{2} \rightarrow e^{-ie^{2}x^{i}} \varphi_{2} \\
    \varphi_{4} \rightarrow e^{-ie^{2}x^{i}} \varphi_{4}
\end{align}
Substituting these into (\ref{cde2t}) and (\ref{cde4t}) gives us
\begin{align}
    i(\tilde{\Sigma}^{i}\tilde{\partial}_{i})\varphi_{2}=-\Sigma^{i}(e^{2} + i\partial_{i})\varphi_{4} + \sqrt{m}\tilde{A}^{0}\varphi_{4} + e^{2} \varphi_{4} \label{cdevc3t}\\
    i(\tilde{\Sigma}^{i}\tilde{\partial}_{i})\varphi_{4}=-\Sigma^{i}(e^{2} + i\partial_{i})\varphi_{2} + \sqrt{m}\tilde{A}^{0}\varphi_{2} - e^{2} \varphi_{2} \label{cdevc4}
\end{align}
Taking the non-relativistic limit, $e^{2}\gg  \sqrt{m}\tilde{A}^{0}$,
\begin{align}
    \sqrt{m}\tilde{A}^{0} \approx 0 \\
    \Sigma^{i}\partial_{i} \approx 0
\end{align}
Hence, (\ref{cdevc4}) becomes
\begin{align}
    i(\tilde{\Sigma}^{i}\tilde{\partial}_{i})\varphi_{4} = -(\Sigma^{i}e^{2}+ e^{2})\varphi_{2} \label{c2devc3}
\end{align}
Substituting (\ref{c2devc3}) in (\ref{cdevc3t}), {we obtain} %MDPI: Please remove the border around the equation if it is not necessary.
\begin{align}
    {\Sigma^{i}(e^{2}+i\partial_{i})\varphi_{4}=-(\tilde{\Sigma}^{i}\tilde{\partial}_{i})^{2}(\Sigma^{i}e^{2} + e^{2})^{-1}\varphi_{4} + \sqrt{m}\tilde{A}^{0}\varphi_{4} + e^{2} \varphi_{4}} \label{2paulieqn6d}
\end{align}
Projecting to the spacetime $\mathcal{M}'_{4}$ with signature $\eta_{\mu \nu} = \text{diag}(-1,-1,-1,+1)$, we impose \mbox{the following}
\begin{align}
    \Sigma^{i} \rightarrow \mathbbm{1} \\
    i\partial_{i} \rightarrow i\partial_{x}
\end{align}
Hence, (\ref{2paulieqn6d}) becomes
\begin{align}
    \mathbbm{1}(e^{2}+i\partial_{x})\varphi_{4}&=-(\tilde{\Sigma}^{i}\tilde{\partial}_{i})^{2}(\mathbbm{1}e^{2} + e^{2})^{-1}\varphi_{4} + \sqrt{m}\tilde{A}^{0}\varphi_{4} + e^{2} \varphi_{4}\\
    i\partial_{x}&=-\frac{1}{2e^{2}}(\tilde{\Sigma}^{i}\tilde{\partial}_{i})^{2} + \sqrt{m}\tilde{A}^{0}
\end{align}
Owing to the resemblance with Pauli's Equation (\ref{paulieqn}), we identify  $\tilde{\Sigma}_{i}$ to be the temporal spin matrices. Now, the spatial spin matrices are associated with \(M_4\) and the temporal spin matrices are associated with \(M_4'\), which gives an interpretation of spin in 6D. 

\section{Proposed Resolution of the EPR Paradox \label{sec4}}

\subsection{Proposal} 
Suppose Alice and Bob are space-like separated inertial observers in our 4D spacetime who are stationary with respect to each other. Both of them have one electron/positron each from a pair in an entangled state. When Alice makes an observation on her electron, the wave function of Bob's positron seems to collapse as if it violates locality. Local hidden variable theories were introduced to understand this, but Bell inequalities show that such theories have an upper bound, and quantum theories violate that bound, showing that local hidden variable theories cannot explain experiments. Quantum theory goes against local realism, and any hidden variable theory must be nonlocal. 

\begin{figure}[H]
  %  \centering
    \includegraphics[width=0.55\textwidth]{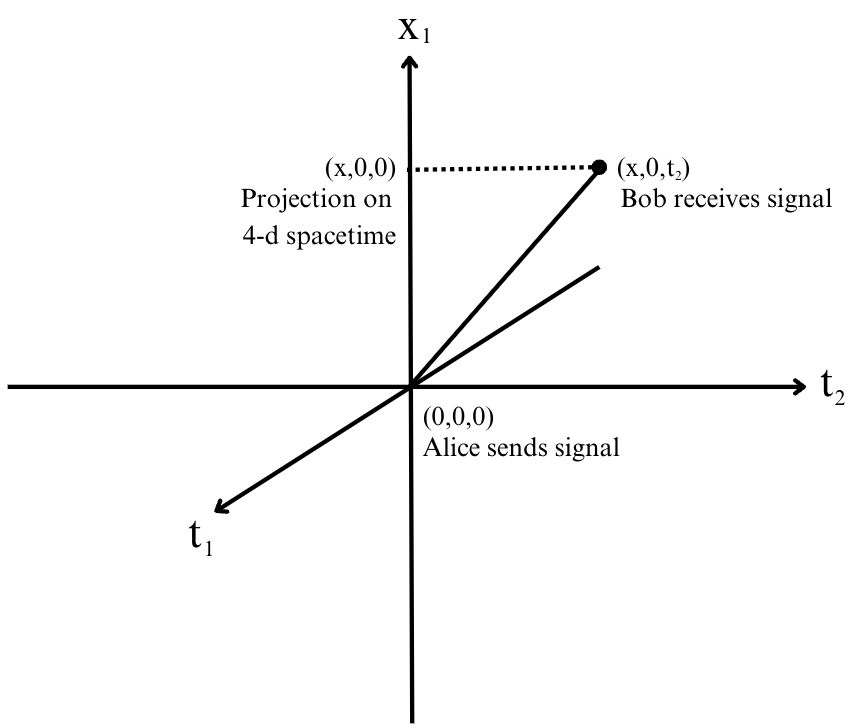} 
    \caption{$t_1$ corresponds to the time axis, which is common between the 6D spacetime and the \mbox{4D spacetime.}}
    \label{spacetime}
\end{figure}

We propose that the nonlocality puzzle can be resolved in the presence of a 6D spacetime of signature $(3,3)$ and by assuming that the electron--positron pair traverses the 6D spacetime, not the 4D spacetime. Consider the 3D illustration of Figure \ref{spacetime}, where $x_1, t_1$ are dimensions in our 4D spacetime and $t_2$ is a time-like dimension in 6D.\vspace{-3pt}

From Figure \ref{spacetime}, the Minkowski spacetime interval can be mathematically expressed as the following line element:
\begin{equation}
ds^2 = dt_2^2 + dt_1^2 - dx_1^2
\end{equation}

Choose \(dt_2\), \(dt_1\), and \(dx_1\) such that the path defined by this line element is inside the light cone of this 3D spacetime \((t_2, t_1, x_1)\), which has the signature \((+, +, -)\), namely, two time-like and one space-like directions. In other words, \(ds^2 > 0\). Now, consider the case that even when \(dt_1 = 0\), the line element is time-like. That is, 
\begin{equation}
ds^2 = dt_2^2 - dx_1^2
\end{equation}
is positive.

Consider the 2D spacetime \((t_1, x_1)\) [which represents our spacetime in which Alice and Bob make measurements]. In such a case, while observing the event, we will naturally assume that \(dt_2 = 0\) and conclude that \(ds^2 = - dx_1^2 < 0\) and that the separation between the sending of a signal and the receiving of the signal is space-like and, hence, nonlocal. This is as though faster-than-light influence has taken place. However, in the 3D universe, the event is local and obeys special relativity.

Thus, the EPR paradox can be resolved if there is an additional time-like direction in the universe. Classical systems do not probe the \(t_2\) direction. Only quantum systems, which obey quantum linear superposition, probe \(t_2\). A photon traveling from Alice to Bob travels along a path that is a quantum superposition of a path in \((t_1, x_1)\) and a path in \((t_2, x_1)\). If we do not know of the photon path in the other submanifold, that lack of knowledge gives rise to the EPR paradox. In the realistic case of 6D spacetime, classical detectors (such as those used by Alice and Bob) exist only in 4D spacetime. These detectors do not probe the two additional time-like dimensions; probing these extra dimensions requires quantum systems confined to the weak length scale.

In essence, we are saying that the path through the other sub-manifold is shorter than the path in ours and takes less time to traverse. The spatial coordinate separation is just \(dx_1\), which is the same in both submanifolds. But the proper physical spatial distance depends on the spacetime metric, and this metric can be different in the two sub-manifolds, making the path in the other sub-manifold shorter in spatial distance, and, hence, it is covered in less time. If \(L_2\) is the proper spatial distance in the other submanifold and \(L_1\) is the proper spatial distance in the usual 4D spacetime, then \(L_2 \ll L_1\). Light takes time \(dt_2 = L_2/c\) to travel from Alice to Bob in the other submanifold. In our 4D spacetime, the amount of \(t_1\) time elapsed while distance \(L_2\) is covered in the other universe is \(L_2/c\), which is far less than the actual travel time \(L_1/c\) in our 4D spacetime. The arrival to Bob appears instantaneous to us, which is why we set \(dt_1 = 0\).

\subsection{Experimental Validation} 
Consider that Alice makes a measurement that causes the electron state to collapse. The signal carrying information about the collapse travels from Alice to Bob  (through $M_4'$) at the speed of light, arriving at the correlated positron after time $t_{1P}=L_2/c$. Clearly, the influence on the positron is not instantaneous but takes a finite time $t_{1P}$ (howsoever small). If Bob makes his measurement on the positron prior to time $t_{1P}$, i.e., prior to the arrival of information of collapse from Alice, evidently, the correlation will not have been established. The results of Bob's measurements in this case will not show a violation of Bell's inequalities. The inequalities will only be violated if Bob's measurements are made later than time $t_{1P}$. In principle, this feature can be tested experimentally.

In practice, however, the experiment is extremely challenging, and essentially impossible, with current technology. This is because in our research program of gravi-weak unification, the pseudo-Riemannian geometry of the manifold $M_4'$ is determined by the weak force, whereas, in our spacetime $M_4$, the cosmic horizon is at about $10^{28}$ cm; in $M_4'$, this same horizon is at the range of the weak force, namely, about $10^{-16}$ cm. In other words, lengths are scaled down by an enormous factor $10^{44}$, and, through $M_4'$, light will travel from our location to the cosmic horizon in merely $10^{-26}$ s, which of course, for all practical purposes, is instantaneous. Therefore, in reality, $t_{1P} < 10^{-26}$ s and a Bell test of the 6D spacetime idea is simply impossible. An observer unaware of the 6D spacetime will infer that the signal traveled through our 4D spacetime at a staggering speed of $10^{44}$c. This is far, far greater than the lower experimental bound of about $10^5$c on the speed of such a correlation signal (assuming that collapse information travels through our 4D spacetime at the speed of light) \cite{Gisinspeed, Cocciaro, Amato}. This experimental bound also tells us that the shrinking of lengths in $M_4'$, relative to $M_4$, is by at least a factor of $10^5$. 

An important caveat in the above reasoning was brought to our attention by \mbox{Gisin \cite{Gisinqm100}.} When the measurements by Alice and Bob are space-like separated, there is no definite causal order (past to future) between Alice's measurement and Bob's measurement; the ordering is frame-dependent. Therefore, such an experiment as the one proposed above, if it could at all be done, would have to be in a universal absolute time, say, in the rest frame of the cosmic microwave background.

The second path through $M_4'$ could be suggestively called a `quantum wormhole'. Our proposal can also be viewed as a rigorous realization of the ER=EPR idea \cite{malda}. Instead of the Einstein--Rosen bridge, we have an extremely short spacetime path connecting Alice and Bob through the second spacetime. It will be interesting to study black hole solutions in 6D spacetime and inquire about how such black holes relate to black hole solutions in $M_4$ and in $M_4'$. 

\section{Physical Motivation for Two Time-like Extra Dimensions \label{sec5}}
There is considerable literature on physics in 
six-dimensional spacetimes with the signature $(3,3)$. The motivations for such considerations are varied, starting 
with desiring as many time dimensions as spatial ones. Quaternions also point to a 6D spacetime, as Lambek explains in his paper titled `Quaternions and three temporal dimensions'  \cite{rf3lambek3}. 
In an insightful paper titled `Germ of a synthesis: spacetime is spinorial, extra dimensions are time-like', Sparling 
\cite{rf3sparling}
observes the relation of null twistor spaces with 6D spacetime. 
Highly relevant for us is also the 1985 paper of Patty and Smalley  \cite{rf3patty} titled `Dirac equation in a six-dimensional spacetime'. The authors show that a (3+3) spacetime can be divided into six copies of (3+1) subspaces. Six-dimensional  spaces are also of interest from the viewpoint of a superluminal extension of (3+1) special relativity, and it has been shown that a 6D spacetime is the smallest one that can accommodate a superluminal as well as a subliminal branch of the (3+1) spacetime \cite{rf3everett}.   Six dimensional = (3+3) spacetimes were studied extensively in a series of papers by Cole \cite{rf3cole} and Teli \cite{rf3teli}. An early work on `quaternions and quantum mechanics' is that by Conway (1948) \cite{rf3conway1948}. Very relevant for us is also the work by Kritov (2021) \cite{rf3kritov}, who shows that the Clifford algebra $Cl(3,0)$ can be used to make two copies of 4D spacetime with relatively flipped signatures. Dartora and \mbox{Cabrera (2009) \cite{rf3dartora}} studied `The Dirac equation in six-dimensional SO(3,3) symmetry group and a non-chiral `electroweak theory''. An old 1950 paper by Podolanski \cite{rf3podo} studies unified field theory in six dimensions, and, in fact, the abstract starts by saying `The geometry of the Dirac equation is actually six-dimensional'. An elegant 2020 paper by Venancio and Batista \cite{rf3batista}  analyzes `Two-Component spinorial formalism using
quaternions for six-dimensional spacetimes'. An insightful 1993 work by Boyling and Cole \cite{rf3boyling} studies the 
six-dimensional (3+3) Dirac equation and shows that particles have spatial spin-1/2 and temporal spin-1/2. See also Brody and Graefe (2011) \cite{rf3brody}. Shtanov and Sahni studied a five-dimensional cosmological model with two time-like
dimensions \cite{sahni}. 
Interestingly, such a universe bounces at high densities and the nature of the bounce
is similar to that in loop quantum cosmology. Insightful studies in 6D spacetime are also due to Pavsic \cite{pavsic1, pavsic2}.

Our interest in 6D spacetime stems from the ongoing research program on unification, known as the $E_8 \otimes E_8$ octonionic theory of unification, recently reviewed by \mbox{Singh (2024) \cite{Singh2024Review}.} This pre-spacetime, pre-quantum theory is a matrix-valued Lagrangian dynamics on a split bi-octonionic space, which obeys the $E_8 \otimes E_8$ unified symmetry prior to the electroweak symmetry breaking. This is also the gravi-weak symmetry breaking, wherein each of the two $E_8$-s branches into four $SU(3)$s. The net result is a 6D spacetime of signature $(3,3)$ broken into two overlapping 4D spacetimes $M_4$ and $M_4'$ and six emergent forces. Four of these are currently known to us and two new forces are predicted, these being $SU(3)_{grav}$ and $U(1)_{DEM}$. On $M_4'$, the breaking $SU(3)\rightarrow SU(2)_L \otimes U(1)_Y\rightarrow U(1)_{em}$ gives rise to the weak force as pseudo-Riemannian geometry of $M_4'$, with the unbroken sector $SU(3)_{color}\otimes U(1)_{em}$ providing the geometry of the vector bundle. The quantum number associated with $U(1)_{em}$ is the electric charge $e$. Correspondingly, on $M_4$,   the breaking of another $SU(3)\rightarrow SU(2)_R \otimes U(1)_{YDEM}\rightarrow U(1)_{DEM}$ gives rise to general relativity as the pseudo-Riemannian geometry of $M_4$, with the unbroken sector $SU(3)_{grav}\otimes U(1)_{DEM}$ providing the geometry of the vector bundle. DEM stands for dark electromagnetism and the quantum number associated with $U(1)_{DEM}$ is the square root of mass $\sqrt m$ (this is the reason for the interchange $e\leftrightarrow \sqrt{m}$ in Section \ref{sec3} when going from $M_4'$ to $M_4$).     

Thus, the weak force is the spacetime geometry of $M_4'$ and general relativity is the spacetime geometry of $M_4$. Both interactions are chiral, the former being left-handed and the latter being right-handed. Together, they are unified into a (non-chiral) gravi-weak symmetry on 6D spacetime, prior to the electroweak symmetry breaking. A detailed analysis leading to these results will be presented in a forthcoming paper \cite{Finster2025}. Thus, in our unification program, it is natural to consider a 6D spacetime with the signature (3,3) as physical reality.

\section{Tsirelson Bound \label{sec6}}

Bell's theorems are a set of closely related results that imply that quantum mechanics is incompatible with local hidden variable theories. Bell's inequality is the statement that when measurements are performed independently on two space-like separated particles in an entangled pair, the assumption that outcomes depend on local hidden variables implies an upper bound on the correlations between the outcomes. As we know, quantum mechanics predicts correlations that violate this upper bound. The CHSH inequality is a particular Bell inequality in which classical correlation (i.e., if local hidden variables exist) can take a maximum value of 2. Quantum mechanics violates this bound, allowing for a higher bound on the correlation, which can take a maximum value of $2\sqrt 2$, known as the Tsirelson bound \cite{Tsirelson}. Popescu and Rohrlich \cite{PR} showed that the assumption of relativistic causality allows for an even higher bound on the CHSH correlation, this value being 4. It is important to ask why the bound coming from causality is higher than the Tsirelson bound. Are there relativistic causal dynamical theories that violate the Tsirelson bound? In a recent paper \cite{Rabsan}, we found this to indeed be the case. We showed that the pre-quantum theory of trace dynamics, from which quantum theory is emergent as a thermodynamic approximation, permits the CHSH correlation to take values higher than $2\sqrt 2$. We interpreted our findings to suggest that quantum theory is approximate and emergent from the more general theory of trace dynamics.

What impact does the extension of spacetime to 6D have on the Tsirelson bound on quantum correlations? Our analysis below suggests that in this scenario as well, the Tsirelson bound could be beaten.

We now start with a general framework where Alice and Bob are each sent a two-state quantum system. Alice can choose to measure one of two quantum observables, $A$ or $A'$. Similarly, Bob can choose to measure $B$ or $B'$. Each of these observables have two possible eigenvalues, $ \pm1$. The entangled system is described by \( \psi_1 \), the state prepared by us, and the superposition state described by \( \psi=\alpha \psi_1 + \beta \psi_2 \) (over the two spacetimes $M_4$ and $M_4'$). Let us find out what happens to the correlations $E(AB)$. The expectation value of the correlation \( E(AB) \) is given by
\begin{equation}
  E(AB) = \langle \psi | A B | \psi \rangle  
\end{equation}

Substituting \( \psi = \alpha \psi_1 + \beta \psi_2 \), we obtain
\begin{equation}
 E(AB) = \langle (\alpha \psi_1 + \beta \psi_2) | A B | (\alpha \psi_1 + \beta \psi_2) \rangle   
\end{equation}
Using the linearity of the inner product and factoring out constants gives
\begin{equation}
    E(AB) = |\alpha|^2 \langle \psi_1 | A B | \psi_1 \rangle + \alpha^* \beta \langle \psi_1 | A  B | \psi_2 \rangle + \beta^* \alpha \langle \psi_2 | A  B | \psi_1 \rangle + |\beta|^2 \langle \psi_2 | A  B | \psi_2 \rangle
\end{equation}
There is an important contribution from cross-terms that may not be zero even if the states are orthogonal. This occurs because the operator AB defined on the complete Hilbert space can rotate \(\psi_2\) in such a way that it positions it at an angle other than \(90\deg\)
  relative to \(\psi_1\). This happens because of the off-block-diagonal terms in \eqref{blockrep}.
  Note that the states are normalized, that is, \( |\alpha|^2 + |\beta|^2 = 1 \). What we can now do is identify the real number coming from the cross-terms as 
\begin{equation}
     r_{AB} = \alpha^* \beta \langle \psi_1 | AB | \psi_2 \rangle  +  \beta^* \alpha \langle \psi_2 | AB | \psi_1 \rangle 
\end{equation}
The CHSH correlation function $F$ is given by
\begin{equation}
   F = E(A, B) + E(A', B) + E(A, B') - E(A', B') 
\end{equation}
This can be expressed as
\begin{equation}
    F = |\alpha|^2 F_1 + |\beta|^2 F_2  + r \quad ;\quad r=r_{AB}+r_{A'B}+r_{AB'}-r_{A'B'}
    \label{rdef}
\end{equation}
where \( F_1 \) and \( F_2 \) are defined as
\begin{equation}
    F_1 = \langle \psi_1 | A B | \psi_1 \rangle + \langle \psi_1 | A' B | \psi_1 \rangle + \langle \psi_1 | A  B' | \psi_1 \rangle - \langle \psi_1 | A'  B' | \psi_1 \rangle
\end{equation}
\begin{equation}
    F_2 = \langle \psi_2 | A  B | \psi_2 \rangle + \langle \psi_2 | A'  B | \psi_2 \rangle + \langle \psi_2 | A  B' | \psi_2 \rangle - \langle \psi_2 | A'  B' | \psi_2 \rangle
\end{equation}
We compute the squares of \( F_1 \) and \( F_2 \):
\begin{equation}
    F_1^2 = \langle \psi_1 | 4I - [A,A'][B,B'] | \psi_1 \rangle
\end{equation}
\begin{equation}
    F_2^2 = \langle \psi_2 |  4I - [A,A'][B,B'] | \psi_2 \rangle
\end{equation}
The square of \( F \), i.e.,  \( F^2 \), in terms of \( F_1 \), \( F_2 \), and $r$ is given by
\begin{equation}
    F^2 = |\alpha|^4 F_1^2 + 2 |\alpha|^2 |\beta|^2 F_1 F_2 + |\beta|^4 F_2^2 + r^2 + 2r(|\alpha|^2F_1+|\beta|^2F_2)
\end{equation}
Now, for the case of the normalized state \( a^2 + b^2 = 1 \), where \( a = |\alpha| \) and \( b = |\beta| \), using the upper bounds \( F_1^2 \leq 8 \) and \( F_2^2 \leq 8 \) and the AM-GM (arithmetic mean--geometric mean) inequality, the expression simplifies to
\begin{equation}
    F^2 \leq 8 (a^4 + b^4) + a^2 b^2 (F_1^2 + F_2^2) + r^2+2r(|\alpha|^2F_1+|\beta|^2F_2)
\end{equation}
which then gives us 
\begin{equation}
    F^2 \leq 8 (a^4 + b^4) + 16a^2 b^2 +r^2+2r(|\alpha|^2F_1+|\beta|^2F_2) \leq 8 + r^2+4\sqrt{2}r
\end{equation}
This shows the possibility that the Tsirelson bound can be violated in some cases, provided that $r> 0$ or $r < -4\sqrt 2$.  If $ -4\sqrt 2 \leq r \leq 0$, the CHSH inequality is obeyed. For $r=4-2\sqrt 2$, the Popescu--Rohrlich bound of 4 on the CHSH correlation $F$ is reached. Curiously, this numerical value $4-2\sqrt 2$ is precisely the gap between the PR value 4 and the Tsirelson bound of $2\sqrt 2$. By defining the variable $z=r+2\sqrt 2$, the contribution of the cross-terms can be suggestively written as $(z^2 - (2\sqrt 2)^2)$, showing $z$ to be the measure of violation of the Tsirelson bound. The bound is obeyed when $|z|\leq 2\sqrt 2$ and violated if $|z|>2\sqrt 2$. The PR bound is equivalent to $|z|=4$. Considering how $r$ was defined in Equation (\ref{rdef}) above, it seems to be the case that these cross-terms are precisely the missing link between the Tsirelson bound and the Popescu--Rohrlich bound. 

It could be that current experiments are unable to detect supra-quantum nonlocal correlations precisely because the consequences of the cross-terms (which make $r$ non-zero) are very hard to detect. This is for the same reasons as those mentioned in Section \ref{sec5}, i.e., the transmission through the path in $M_4'$ is enormously quick. 

Also, the challenge for appropriate experiments 
is that whichever observables we currently measure act only on $\psi_1$; they never
rotate $\psi_1$ outside of $M_4$, at least, we cannot claim this
theoretically. For the observables that complete such rotations, one might wonder if they correspond to any physical quantity. Our assessment is that such observables will be physical; they likely relate to the weak interaction since the weak interaction determines the spacetime geometry of $M_4'$. Therefore, it could be that high energy experiments proposed to test Bell's inequalities at colliders \cite{Barr} might be the ideal place to look for violations of the Tsirelson bound because such experiments are likely to be sensitive to the weak scale. The quantum field that transmits the signal through $M_4'$ could be dark electromagnetism, the $U(1)_{DEM}$ gauge field mentioned in the previous section. Its carrier is the massless dark photon that couples to the square root of the fermionic mass.

Considering that the weak force is the spacetime geometry of $M_4'$, there ought to exist weak waves, analogous to gravitational waves and electromagnetic waves, but with wavelengths smaller than $10^{-16}$ cm. If ever such waves are detected in experiments, they could be a possible indicator of 6D spacetime.

The presence of three times sometimes raises inquiries as to their physical implications. For instance, which of these is the time that flows and, hence, defines an arrow of time? Our stance is that these three times are time {\it coordinates}, on the same footing as the three spatial coordinates. They are mechanistic and reversible and cannot by themselves provide a time arrow. The role of fundamental time is played by a parameter that, by itself, is not part of the spacetime manifold. In our research program, such a parameter is the 
so-called Connes time in non-commutative geometry, whose origin lies in the Tomita--Takesaki theorem for von Neumann algebras \cite{Connes}. 

\section{Critique \label{sec7}}
In this section, we address several critical questions a reader could raise regarding the assumptions underlying our proposal for resolving the EPR paradox. 

\begin{itemize}
\item What is the underlying theory of unification that motivates the 6D spacetime with the signature $(3,3)$?

The starting point for the proposed unification theory is quantum foundational. We seek a reformulation of quantum field theory, at all energy scales, that does not depend on a background classical time. This new theory, which is reviewed 
in \cite{singhdice2022}, is a pre-quantum, pre-spacetime theory. The dynamics is described by Adler's theory of trace dynamics \cite{Adler}, which is a matrix-valued Lagrangian dynamics. On time scales larger than Planck time, 
conventional quantum field theory is emergent from trace dynamics in a coarse-grained approximation.

Given a Riemannian geometry, the eigenvalues 
of the Dirac operator on the spacetime manifold are treated as classical dynamical variables \cite{spectral, landi}. In trace dynamics, these eigenvalues are raised to the status of an operator, the very Dirac operator of which they are eigenvalues in fact. Each spacetime point is replaced by a non-commuting number, the 16D split bioctonion, and the symmetry group of the theory is assumed to be $E_8 \otimes E_8$, as explained in \cite{Kaushik, Singh2024Review}. General relativity and the standard model forces emerge from this unified theory after the electroweak symmetry breaking. We also thus explain how the 6D spacetime with two embedded 4D spacetimes (of relatively flipped signature) 
arises. The two 4D spacetimes have one space and one time direction in common, but each of the spacetimes has its own distinct metric. The Riemannian geometry of our 4D spacetime is described by the general theory of relativity, whereas the geometry of the other 4D spacetime is a description of the weak interaction as an external symmetry \cite{Finster2025}. From our vantage point, the two additional time-like directions can also be thought of as internal symmetry directions, as required by the gauge-symmetric description of the weak force. Prior to the electro-weak symmetry breaking, we have gravi-weak unification on a 6D spacetime. The theory has been used to attempt a derivation of some of the 
fundamental constants of the standard model \cite{mass, ckm}.

\item We assume that detection devices are classical and live in our 4D spacetime. Why should that be so?

This is an entirely valid inquiry and one that requires a physical process to explain the emergence of our 4D classical spacetime from the underlying 6D spacetime. In our theory, the electroweak symmetry breaking is also a quantum-to-classical transition. It is this transition that effectively localizes classical systems to 4D because the remaining two time dimensions are compactified to the weak interaction time scale of about $10^{-18}$ m/c$\sim$$10^{-26}$ s. This is a fundamental time scale in the other 4D spacetime, where it replaces Planck time, which is $\sim$$10^{-43}$ s. The other 4D spacetime acquires its own distinct metric where the basic length scale is the scale of the weak interaction $10^{-18}$ m. Detectors, which are by definition classical and macroscopic, effectively reside in our 4D spacetime because they do not experience the weak force. On the other hand, quantum systems experience the universal weak force and, hence, effectively reside in all six dimensions.

\item What is the physical mechanism that quickly transmits the information  of collapse from Alice to Bob through the second 4D spacetime?

The octonionic theory predicts a new long range force, a $U(1)$ gauge symmetry dubbed dark electromagnetism, which couples to the square-root of mass. The associated massless gauge boson, named the dark photon, resides in the
second 4D spacetime and is a promising candidate for locally transmitting information about collapse from Alice to Bob. Another possible candidate for transmission are the weak waves alluded to above. Both these, the dark photon
as well as weak waves, are experimentally falsifiable predictions of our unification theory.

As for collapse of the wave function, it is a dynamical process known as continuous spontaneous localization [CSL, objective collapse models] \cite{GRW}. This is a well-known proposed generalization of quantum theory. Dynamical 
collapse is an inevitable consequence of trace dynamics; moreover, in \cite{Kakade}, we demonstrate how random dynamical collapse arises from the underlying dynamical theory. Once again, this is a falsifiable prediction of our 
unification proposal.

\item What is the justification for our assumption of an absolute time, which is outside the realm of standard special relativity?

Einstein gave up on Newton's absolute time when making the transition from Galilean relativity to special relativity. However, it is useful to recall the Lorentz aether theory, an equivalent formulation of special relativity that
retains an absolute frame. Lorentz argued in 1913 to the effect  that  there is no significant difference between his theory and the rejection of a preferred reference frame, as in the theory of Einstein and Minkowski; thus,  it is a matter of taste as to which theory one prefers \cite{wikilorenz}. This viewpoint was supported in later years by other researchers \cite{recent} who noted that both the theories make the same experimental predictions \cite{Janssen}. Thus, we note that even at the level of special relativity, an absolute time [in which the 4D spacetime manifold evolves] is admissible, though it is not strictly necessary. On the other hand, an absolute time becomes essential and unavoidable  in a truly relativistic quantum theory, in which space and time ought to be treated on par. Not only should position be treated as an operator but time as well. This necessitates that an additional external 
parameter (which is not an operator) be introduced so as to keep track of evolution. Such a relativistic formulation of quantum mechanics does exist; it is the Stueckelberg--Horowitz formulation \cite{Horowitz}, even though the external time parameter is introduced in an ad hoc manner. Conventional quantum mechanics, in which time is not an operator, is recovered as an approximation to the more general theory, when spontaneous localization in time transforms operator time to classical parameter time. 

When considering multiple time dimensions, as in the present paper, one is confronted with a paradoxical situation: we experience only one time dimension; this is the time that flows. How to reconcile this with having three times? This 
paradox is resolved in two stages. Firstly, an absolute time parameter is introduced, which is distinct from these three times. These three times, while fundamentally operator-valued in nature, are approximately  treated as 
classical in the present paper. Secondly, two of these time dimensions are compactified to the scale of the weak force (i.e., $10^{-26}$ s) and are hence not perceived in the macroscopic classical world. Therefore, the third time dimension (which is a part of our 4D spacetime) is effectively the only one that remains in the classical world and can therefore be identified, without loss of generality,  with absolute time.

In our theory, absolute time also arises in a natural manner and is not ad hoc. The geometry of the pre-quantum, pre-spacetime theory is a non-commutative geometry. By virtue of its non-commutativity, the theory admits a one-parameter family of outer automorphisms; these serve to play the role of a time parameter, as emphasized by Connes \cite{connes1, connes2}. We hence call it Connes time; for further application of this time, see also Connes and 
Rovelli \cite{Connes}. 

\item What is the justification for asserting that in the other 4D spacetime, the distance between Alice and Bob is much less than in our 4D spacetime?

It can be argued that the extent of distances in the second 4D spacetime is of the order $L^{1/3}$, where $L$ is the physical extent of our 4D spacetime, measured in Planck length units. This claim is based on the so-called
holographic length uncertainty relations~\cite{singhholo}. Thus, $L$ is of the order $10^{28}/10^{-33}$$\sim$$10^{61}$ Planck units. If follows that maximum distances in the second 4D spacetime are of the order $10^{20}$ Planck units and, hence, about $10^{-13}$ cm. This is not very far from the weak interaction length scale of $10^{-16}$~cm, which sets the scale of distance in the other 4D spacetime. In other words, we are saying that the cosmological horizon, which is at $10^{28}$ cm 
in our spacetime, is only some $10^{-16}$ cm away in the other 4D spacetime.

\item In our framework, is the influence of collapse instantaneous?

No. The collapse information travels from Alice to Bob at the speed of light, so that there is a finite time interval during which Bob does not yet experience the influence of collapse at Alice's end. But this finite time
interval is extremely small, being of the order of $10^{-26}$ s or smaller. This is not ruled out by current Bell experiments~\cite{Gisinspeed, Cocciaro, Amato}.

When we talk of a superposition of two states of the mediating field in the two 4D universes, we do not mean it as a quantum superposition with respective probability amplitudes in the two branches. [However, the entangled EPR pair of particles (but not the mediating field) is in quantum superposition across the two 4D spacetimes, which we explain
later.] What we mean is that when wave function collapse takes place at Alice’s end, the information about the collapse is transmitted independently through two distinct 4D spacetimes by {\bf {two distinct mediating fields}%MDPI: Please confirm if the bold should be retained.
}. Note that because the (classical) detectors are exclusively in our 4D spacetime, collapse of the EPR pair happens necessarily onto our 4D spacetime. Yet, information about collapse travels through both 4D spacetimes separately. In our spacetime, the transmitting field is light (the photon). In the other 4D spacetime, the transmitting field is dark electromagnetism (dark photon) also moving at the speed of light. The dark photon invariably arrives at Bob’s end before the ordinary photon does, causing full wave function collapse in the entangled particle pair; hence, the ordinary photon is never observed at Bob’s end. The only way to observe the photon at Bob’s end is to have the second particle of the entangled pair continue in a superposed state of alternatives, but that state has already collapsed before the ordinary photon arrives at Bob’s end. If the entangled pair does not have electromagnetic interaction, there remains the possibility that in our 4D spacetime, the mediating field is gravitational waves (because gravity is universal), and these are not detected at Bob’s end because of the extremely feeble nature of gravity. 

When we talk about the state $\psi = a |U_1\rangle + b |U_2\rangle$ (an element of the complete Hilbert space $H+H’$) of the entangled EPR pair as a superposition across the two 4D spacetimes (which we have analyzed and justified for Dirac particles using the 6D Dirac equation), the measurements on this state necessarily lead to a state in $H$ (Hilbert space associated with our 4D spacetime) with probability one because the observables that we measure using classical detectors (confined to our 4D spacetime) are associated with $H$ and their eigenstates only span $H$. We note that the state $\psi$ cannot be written as a linear combination of eigenstates of any observable in $H$, and the standard Born rule for probability calculations cannot be directly applied in this case. Usually in quantum mechanics, we encounter Hilbert spaces and their tensor products with the observable eigenstates spanning them, but, here, in this case, it is different. The standard Born rule is applied for $|U_1\rangle$ (an element of $H$). However, the mediating field carrying information of this collapse goes through both the 4D spacetimes. 

\item  We start from quantum theory in 6D spacetime and take the classical limit to arrive at a 4D spacetime. Is not this procedure arbitrary? It seems to break the  general principle of covariance according to which the classical
limit of quantum theory should be coordinate-independent. What happened to the other two times, while descending from 6D to 4D? How does one dimension of time get preferably selected for \mbox{our universe?}

The transition from 6D spacetime to 4D spacetime is a consequence of spontaneous symmetry breaking, this being the breaking of electroweak symmetry, which is also a quantum-to-classical transition \cite{Kaushik, Finster2025}. The 6D spacetime 
branches into two overlapping 4D spacetimes; ours arises from the breaking of an $SU(2)_R \otimes U(1)_{Y'}$ symmetry, giving rise to general relativity as the geometry of 4D spacetime. The other 4D spacetime arises 
from the breaking of the $SU(2)_R \otimes U(1)_Y$ symmetry, and the weak force determines the geometry of this 4D spacetime. Dimensions in this latter spacetime are compact and of the order of the weak length/time scale. 
In this sense, from the perspective of our 4D spacetime, the two additional time dimensions are macroscopically imperceptible. The general principle of covariance is obeyed in both the 4D spacetimes separately
because the symmetry breaking dissociates one 4D spacetime from the other 4D spacetime: only our 4D spacetime is classical,  the other 4D spacetime is not.

We can explain this also in a slightly different manner. On the 6D spacetime, assumed to be Minkowski and with a signature of $(3,3)$, we have an $SU(3) \otimes SU(3)$ Yang--Mills gauge theory. It is important to note that, here, on 6D, we are not talking of gravity and general covariance; we only have Lorentz invariance and Yang--Mills fields. At the epoch of electroweak symmetry breaking, the first $SU(3)$  branches as $SU(3) \rightarrow SU(2)_L \otimes U(1)_Y \rightarrow U(1)_{em}$, giving rise to the weak force and the unbroken symmetry of electromagnetism. The second $SU(3)$ branches as $SU(3) \rightarrow SU(2)_R \otimes  U(1)_{Y'} \rightarrow  U(1)_{DEM}$, where the unbroken symmetry $U(1)_{YDEM}$ is dark electromagnetism. Concurrently, the 6D spacetime branches into two overlapping copies of 4D spacetimes, each of which has its own Minkowski metric and its own light cone. The first 4D spacetime has three space-like directions and one time-like direction and the signature $(+, -, -, -)$. The second 4D spacetime has three time-like directions and one space-like direction and the signature $(+, +, +, -)$. If we denote the coordinates of the 6D spacetime as 
$(t_3, t_2, t_1, x_1, x_2, x_3)$, then the first 4D spacetime has coordinates $(t_1, x_1, x_2, x_3)$ and the second 4D spacetime has coordinates $(t_3, t_2, t_1, x_1)$.  There is a spontaneous symmetry breaking of $SO(3,3)$ into two 
non-commuting copies of $SO(1,3)$ and $SO(3,1)$.
Our results are consistent with those of Patty and Smalley [`{\it Dirac equation in a six-dimensional spacetime: Temporal polarization for {subluminal interactions}%MDPI: Please confirm if the italics should be retained.
}'] \cite{rf3patty}, who note in their abstract that '\ldots the electromagnetic field introduces a polarization of the temporal axes and that this polarization effect divides the (3+3) spacetime into six (3+1) Lorentzian subspaces. Subluminal interactions which involve fields and particles within a specified (3+1) subspace do not introduce multitemporal motion'. Thus, one is justified in considering Lorentz invariance in the first 4D spacetime distinct from Lorentz invariance in the second 4D spacetime. [We are only investigating two out of the six (3+1) subspaces; the remaining four are left aside for a future study]. On the first 4D spacetime, the $SU(2)_R$ gauge symmetry is complexified to $SL(2,C)$, and its gauging is what gives rise to the general theory of relativity; the theory can also be
equivalently thought of as general covariance of the 4D spacetime. On the second 4D spacetime, the $SU(2)_L$ gauge symmetry is complexified to another $Sl(2,C)$ copy, and its gauging can be interpreted as a Riemannian geometry 
interpretation of the weak force (on scales smaller than the range of the weak force). The two 4D spacetimes have their own respective 4D metrics, as also implied by Patty and Smalley \cite{rf3patty}.

\end{itemize}

%\newpage

\section{Conclusions \label{sec8}}
We demonstrated a possible resolution of the EPR paradox by adding extra time-like coordinates. This required the quantum system to be in superposition across \(M_4\) and \(M
_4'\), and this was shown for spinors using the Dirac equation in 6D. We constructed the 6D Dirac equation using the split biquaternions and analyzed how this reduces to the Dirac equation in \(M_4\) and \(M_4'\). It was also crucial for the argument to interpret spin in both these submanifolds. This was achieved by associating Pauli spin matrices to both these submanifolds using the non-relativistic limit of the Dirac equation to obtain Pauli equations for \(M_4\) and \(M_4'\). An experimental validation was also proposed and it was shown that the Tsirelson bound can be violated in some cases because of the presence of interference terms arising when the 6D spacetime is decomposed into two 4D spacetimes and additional operators on the Hilbert space that rotate \(\psi_{M_4}\) such that it leaves a component in \(\psi_{M_4'}\).

After the first version of this paper was brought out as a preprint in early February 2025, we became aware of several other related papers. These papers also attempt to resolve the quantum nonlocality puzzle by adding extra dimensions or a topological structure to our 4D spacetime. The most significant among these articles is the work of Pettini \cite{Pettini}, which predates ours, and, to our knowledge, it is the first paper to suggest additional temporal dimensions as a resolution of the EPR paradox. Our work differs from Pettini's in that we motivate the additional temporal dimensions through our program on gravi-weak unification. We also justify how these additional time dimensions help beat the Tsirelson bound. In another work, physics with multiple time dimensions, including the initial value problem,  was considered in some detail by Weinstein \cite{Weinstein2008}. Genovese~\cite{Genovese} examined how extra spatial dimensions might help resolve the EPR paradox. In a very interesting investigation, Kauffman \cite {Kauffman} considered a topological enhancement of 4D spacetime via a quantum tensor network to arrive at an Einstein--Rosen bridge-like structure and formalize the ER=EPR proposal. It is very interesting to question whether, for a large tensor network, the extension could approximate a smooth manifold structure and an embedding of 4D spacetime in a higher spacetime of the kind considered by us (we thank Louis Kauffman for suggesting this possibility). 
%%%%%%%%%%%%%%%%%%%%%%%%%%%%%%%%%%%%%%%%%%

\vspace{6pt} 
\authorcontributions{MF contributed  to the conceptualization, methodology, validation, and writing of {this article}. TPS contributed  to the conceptualization, methodology, validation, and writing of {this article}. PSW contributed  to the conceptualization, methodology, validation, and writing of {this article}.%MDPI: Please try to revise this section according to the template.
 %MDPI: For research articles with several authors, a short paragraph specifying their individual contributions must be provided. The following statements should be used ``Conceptualization, X.X. and Y.Y.; methodology, X.X.; software, X.X.; validation, X.X., Y.Y. and Z.Z.; formal analysis, X.X.; investigation, X.X.; resources, X.X.; data curation, X.X.; writing---original draft preparation, X.X.; writing---review and editing, X.X.; visualization, X.X.; supervision, X.X.; project administration, X.X.; funding acquisition, Y.Y. All authors have read and agreed to the published version of the manuscript.'', please turn to the  \href{http://img.mdpi.org/data/contributor-role-instruction.pdf}{CRediT taxonomy} for the term explanation. Authorship must be limited to those who have contributed substantially to the work~reported.
All authors have read and agreed to the published version of the manuscript.
}

\funding{This research received no external funding.}

\dataavailability{No new data were created or analyzed in this study. Data sharing is not applicable to this article.}

\acknowledgments{For participation in the early stages of this work, we would like to thank Arpit Chhabra, Yash Gupta, Satwik Mittal, Abhijeet Mohanty, Navaneeth, and Aayush Srivastav. For helpful discussions, it is a pleasure to thank Priyanka Giri, Nicolas Gisin, Jose Isidro, Louis H Kauffman, Nehal Mittal, Hendrik Ulbricht, and Harald Weinfurter.}

\conflictsofinterest{The authors declare no conflict of interest.}  %MDPI: Declare conflicts of interest or state ``The authors declare no conflicts of interest.'' Authors must identify and declare any personal circumstances or interest that may be perceived as inappropriately influencing the representation or interpretation of reported research results. Any role of the funders in the design of the study; in the collection, analyses or interpretation of data; in the writing of the manuscript; or in the decision to publish the results must be declared in this section. If there is no role, please state ``The funders had no role in the design of the study; in the collection, analyses, or interpretation of data; in the writing of the manuscript; or in the decision to publish the results''.

\begin{adjustwidth}{-\extralength}{0cm}

\reftitle{References}

\PublishersNote{}
%\isPreprints{}{% This command is only used for ``preprints''.
\end{adjustwidth}
\end{document}